\documentclass[pdflatex,sn-mathphys-num]{sn-jnl}

\usepackage{mathrsfs}
\usepackage[T1]{fontenc}
\usepackage[utf8]{inputenc}
\usepackage{graphicx}
\usepackage{multirow}
\usepackage{amsmath,amssymb,amsfonts}
\usepackage{amsthm}
\usepackage[title]{appendix}
\usepackage{xcolor}
\usepackage{textcomp}
\usepackage{manyfoot}
\usepackage{booktabs}
\usepackage{algorithm}
\usepackage{algorithmicx}
\usepackage{algpseudocode}
\usepackage{listings}
\usepackage{ulem}

\newcommand{\ket}[1]{\left|\, #1 \,\right\rangle}

\allowdisplaybreaks

\begin{document}

\title[Eigenvalue-based Richardson spectra]{\sf\Large 
Eigenvalue-based computation of Richardson pairing spectra and finite-size thermodynamics}

\author*[1,2]{\fnm{Grzegorz} \sur{Biskowski}}\email{grzegorz.biskowski@usz.edu.pl}

\author*[2]{\fnm{Marcin R.} \sur{Pi\k{a}tek}}\email{marcin.piatek@usz.edu.pl}

\author[2]{\fnm{Franco} \sur{Ferrari}}\email{franco.ferrari@usz.edu.pl}

\affil[1]{\normalsize\orgdiv{Doctoral School}, \orgname{University of Szczecin}, 
\orgaddress{\street{Adama Mickiewicza 16}, \city{Szczecin}, 
\postcode{70--384}, \country{Poland}}}

\affil[2]{\normalsize\orgdiv{Institute of Physics}, \orgname{University of Szczecin}, 
\orgaddress{\street{Wielkopolska 15}, \city{Szczecin}, 
\postcode{70--451}, \country{Poland}}}

\abstract{We present a numerical workflow for the reduced BCS Richardson model, combining the eigenvalue-based Bethe equations, 
Baxter-polynomial reconstruction, and complex root finding. Newton--Raphson iteration solves the nonlinear equations for the 
eigenvalue-based variables; a linear system yields the polynomial coefficients, and Laguerre iteration with polynomial deflation 
extracts the rapidities. We apply the procedure to equally spaced, quadratic, and hydrogen-like non-degenerate single-particle 
spectra as the pairing strength varies. For finite systems, we calculate canonical observables from the complete seniority-zero 
spectrum and check benchmark spectra against exact diagonalization. In an eight-level, four-pair system with a hydrogen-like 
spectrum, the heat capacity has two maxima at weak coupling. With increasing coupling, the low-temperature maximum shifts upward 
and the high-temperature maximum downward; the latter ceases to be a distinct local maximum near $g\simeq0.03515$ in the units 
used. The equally spaced benchmark instead has a single dominant maximum over the coupling interval displayed for that spectrum. 
This comparison illustrates how single-particle level spacings can shape finite-size thermal response. The C++ code is publicly 
available; numerical instabilities remain possible near ill-conditioned configurations and at larger sizes.}


\maketitle

\section{Introduction}\label{sec1}
Exactly solvable pairing models provide a setting in which interacting many-body spectra and thermodynamic observables can be 
computed without relying on a mean-field approximation. The reduced Bardeen--Cooper--Schrieffer (BCS) pairing Hamiltonian is a 
central example: its eigenstates can be expressed through Richardson's Bethe ansatz, with the many-body energies determined by a 
set of rapidities satisfying coupled nonlinear equations~\cite{Richardson1963,Richardson1964,Dukelsky:2004re}. The model is part 
of the wider Richardson--Gaudin family of quantum integrable systems, whose algebraic structure and applications have been 
reviewed extensively~\cite{Dukelsky:2004re,Claeys:2018zwo}.

Computing the rapidities numerically is challenging because the Richardson equations contain terms that become singular when 
rapidities approach one another or a pair level. Along a coupling scan, real rapidities can coalesce and continue as complex-
conjugate pairs. These features make continuation, polynomial reconstruction, and root extraction important parts of a practical 
numerical implementation. In the present work we restrict the numerical examples to non-degenerate single-particle spectra, so 
that all input pair energies are distinct.

Here we assemble these ingredients into a transparent C++ workflow for the reduced BCS Richardson model. The approach uses the 
established eigenvalue-based formulation~\cite{Faribault2011}: nonlinear equations for the eigenvalue-based variables are solved 
by a Newton--Raphson procedure, the Baxter polynomial is reconstructed from a linear system, and its roots are obtained using the 
Laguerre method with polynomial deflation. We restrict the present implementation and its examples to non-degenerate input 
spectra. The aim is not to introduce a new integrable model or a new Bethe-ansatz formalism, but to document a computational 
procedure and test it across representative non-degenerate spectra.

We also use the resulting Richardson spectrum to evaluate canonical thermodynamic observables in finite systems at fixed particle 
number. The main thermal example is an eight-level system containing four pairs with a hydrogen-like single-particle spectrum. 
Its heat capacity displays two maxima at weak coupling; the low-temperature maximum moves upward and the high-temperature maximum 
moves downward as the interaction grows, until the higher-temperature maximum ceases to be a local maximum near $g\simeq0.03515$ 
in the chosen units. The equally spaced and quadratic benchmarks instead exhibit a single dominant maximum over the coupling 
ranges displayed for those spectra. These calculations provide a concrete illustration of how the distribution of single-particle 
levels can influence finite-size thermal response. Since the study covers a limited set of finite spectra, we treat this 
behaviour as a benchmark result rather than as a universal property of Richardson systems.

The paper is organized as follows. Section~\ref{Ch2} introduces the reduced BCS Hamiltonian, its Richardson--Gaudin structure, 
and the Yang--Yang formulation. Section~\ref{Ch3} presents the eigenvalue-based equations and the reconstruction of rapidities 
for non-degenerate spectra. Section~\ref{Ch4} describes the numerical implementation. Section~\ref{Ch5} reports the spectral 
examples and the finite-temperature results, and the final discussion summarizes the findings, limitations, and directions for 
further work.

\section{Richardson--Gaudin models and Bethe equations}
\label{Ch2}
\subsection{The Richardson reduced BCS model}
We consider the reduced BCS Hamiltonian
\begin{equation}
	H_{\rm rBCS}
	=
	\sum_{j,\sigma}\epsilon_j c_{j\sigma}^{\dagger}c_{j\sigma}
	-g\sum_{i,j}c_{i+}^{\dagger}c_{i-}^{\dagger}c_{j-}c_{j+},
	\label{Hfull}
\end{equation}
where $\epsilon_j$ are the single-particle energies, $g>0$ is the
pairing strength, and $\sigma=\pm$ labels time-reversed states.
The reduced interaction scatters pairs of time-reversed particles
between the active single-particle levels.

For a fixed number of fermions, \(N=2M+K\),
we distinguish $M$ Cooper pairs and $K$ blocked, singly occupied
fermion states. Let $\mathcal U$ denote the set of levels available for pair
scattering and $\mathcal B$ the blocked set. The blocked particles do
not participate in the pairing interaction, and their contribution to
the energy is
\begin{equation}
	E_{\mathcal B}=\sum_{i\in\mathcal B}\epsilon_i.
\end{equation}
The nontrivial interacting problem is therefore the paired sector.

Introducing hard-core pair operators,
\begin{equation}
	b_j^\dagger=c_{j+}^\dagger c_{j-}^\dagger,
	\qquad
	b_j=c_{j-}c_{j+},
\end{equation}
with
\begin{equation}
	[b_i,b_j^\dagger]
	=
	\delta_{ij}\left(1-2b_i^\dagger b_i\right),
	\qquad
	[b_i^\dagger b_i,b_j^\dagger]
	=
	\delta_{ij}b_j^\dagger,
\end{equation}
the Hamiltonian restricted to the paired subspace becomes
\begin{equation}
	\label{HU}
	H_{\mathcal U}
	=
	\sum_{j\in\mathcal U}2\epsilon_j b_j^\dagger b_j
	-g\sum_{i,j\in\mathcal U}b_i^\dagger b_j.
\end{equation}
The energy of a pair at vanishing coupling is therefore $2\epsilon_j$.

The exact eigenstates of \eqref{HU} can be written in Richardson's
Bethe-ansatz form
\begin{equation}
	\ket{\Psi_M}_{\mathcal U}
	=
	\prod_{\nu=1}^{M}B_\nu^\dagger\ket{0},
	\qquad
	B_\nu^\dagger
	=
	\sum_{j\in\mathcal U}
	\frac{b_j^\dagger}{2\epsilon_j-E_\nu},
	\label{BA}
\end{equation}
where the complex parameters $E_\nu$ are the Richardson rapidities.
The energy of the paired sector is
\begin{equation}
	E_{\mathcal U}=\sum_{\nu=1}^{M}E_\nu,
	\label{Epair}
\end{equation}
so that the total energy, including blocked particles, is
\begin{equation}
	E=E_{\mathcal B}+E_{\mathcal U}.
	\label{Etot}
\end{equation}

Substitution of the ansatz \eqref{BA} into the Schr\"odinger equation
gives
\begin{align}
	H_{\mathcal U}\ket{\Psi_M}_{\mathcal U}
	&=
	E_{\mathcal U}\ket{\Psi_M}_{\mathcal U}
	\nonumber\\
	&\quad+
	\sum_{\nu=1}^{M}
	\left[
	1
	-g\sum_{j\in\mathcal U}\frac{1}{2\epsilon_j-E_\nu}
	+2g\sum_{\substack{\mu=1\\\mu\neq\nu}}^{M}
	\frac{1}{E_\mu-E_\nu}
	\right]
	B_0^\dagger
	\prod_{\substack{\eta=1\\\eta\neq\nu}}^{M}
	B_\eta^\dagger\ket{0},
	\label{Prop1}
\end{align}
where
\begin{equation}
	B_0^\dagger=\sum_{j\in\mathcal U}b_j^\dagger.
\end{equation}
The unwanted terms vanish when the rapidities satisfy the
Richardson equations,
\begin{equation}
	\boxed{
		\frac{1}{g}
		-\sum_{j\in\mathcal U}\frac{1}{2\epsilon_j-E_\nu}
		+2\sum_{\substack{\mu=1\\\mu\neq\nu}}^{M}
		\frac{1}{E_\mu-E_\nu}
		=0,
		\qquad
		\nu=1,\ldots,M.
	}
	\label{REq}
\end{equation}
These equations constitute the nonlinear core of the exact solution.
In the present work the active pair levels are taken to be non-degenerate;
blocked orbitals, when considered, are excluded from the pairing sums.

\subsection{Richardson--Gaudin structure}

The integrability of the reduced BCS Hamiltonian is connected to the
rational $\mathfrak{su}(2)$ Gaudin model~\cite{Gaudin1976,Gaudin2014,Dukelsky:2004re}. Introducing pseudospin
operators
\begin{equation}
	S_j^+=b_j^\dagger,
	\qquad
	S_j^-=b_j,
	\qquad
	S_j^z=\frac12-b_j^\dagger b_j,
\end{equation}
the pairing problem is embedded into the commuting family of rational
Gaudin Hamiltonians. In one common convention they are
\begin{equation}
	\label{Gaudin}
	H_{{\rm G},i}
	=
	\sum_{j\neq i}
	\frac{
		S_i^zS_j^z
		+\frac12\left(S_i^+S_j^-+S_i^-S_j^+\right)
	}{\epsilon_i-\epsilon_j}.
\end{equation}
The corresponding Richardson--Gaudin conserved charges can be chosen
as
\begin{equation}
	\label{Ri}
	R_i
	=
	-S_i^z
	-g
	\sum_{j\neq i}
	\frac{\mathbf S_i\cdot\mathbf S_j}{\epsilon_i-\epsilon_j},
\end{equation}
with convention-dependent overall normalizations of $g$ and the
Gaudin operators. The important point for the present work is that
the $R_i$ form a mutually commuting family and that the Richardson
rapidities provide their common Bethe ansatz spectral data.

We will not use the commuting charges directly in the numerical
implementation below. Their role here is to identify the Richardson
model as a member of the rational Richardson--Gaudin class and to
motivate the use of Bethe ansatz variables rather than a direct
diagonalization of the many-body Hamiltonian.

\subsection{Yang--Yang formulation}
\label{YY}
The Richardson equations also arise as stationarity conditions of a
Yang--Yang function. Defining \(z_i=2\epsilon_i\), we introduce \cite{Sierra:1999mp}
\begin{align}
	\mathcal W^{\rm R}(\mathbf z,\mathbf E)
	={}&
	-\sum_{i<j}^{L}\log(z_i-z_j)
	-4\sum_{\nu<\mu}^{M}\log(E_\nu-E_\mu)
	\nonumber\\
	&\quad
	+2\sum_{i=1}^{L}\sum_{\nu=1}^{M}\log(z_i-E_\nu)
	+\frac{1}{g}
	\left(
	-\sum_{i=1}^{L}z_i
	+2\sum_{\nu=1}^{M}E_\nu
	\right).
	\label{U}
\end{align}
Indeed,
\begin{equation}
	\frac{\partial\mathcal W^{\rm R}}
	{\partial E_\nu}
	=
	2\left[
	\frac{1}{g}
	-\sum_{i=1}^{L}\frac{1}{z_i-E_\nu}
	+2\sum_{\substack{\mu=1\\\mu\neq\nu}}^{M}
	\frac{1}{E_\mu-E_\nu}
	\right],
\end{equation}
so that
\begin{equation}
	\frac{\partial\mathcal W^{\rm R}}
	{\partial E_\nu}=0
	\qquad
	\Longleftrightarrow
	\qquad
	\text{Richardson equation \eqref{REq}}.
\end{equation}
This variational formulation provides an alternative interpretation of the rapidities as interacting degrees of freedom in an 
effective electrostatic problem. Although the numerical method developed below does not require direct minimization or 
stationary-point searches of $\mathcal W^{\rm R}$, the Yang--Yang formulation is useful for understanding the structure of the 
equations and for identifying possible future numerical as well as analytical strategies.

There is also a useful connection with conformal field theory. The Richardson solution of the reduced BCS model admits a 
free-field realization in terms of suitable perturbed and unperturbed $\widehat{\mathfrak{su}}(2)_k$ 
Wess--Zumino--Witten (WZW) conformal blocks~\cite{Sierra:1999mp,Sedrakyan:2010}. 
A related relation between Coulomb-gas representations, Dotsenko--Fateev saddle-point actions, 
classical Virasoro conformal blocks, and Richardson--Gaudin Bethe equations was developed in~\cite{PNPP}. These constructions 
motivate a complementary analytical programme based on conformal blocks, alongside the numerical approach pursued here. We do not 
use CFT methods in the calculations reported in this paper; their possible extensions, including irregular conformal blocks and 
irregular Knizhnik--Zamolodchikov systems, are briefly discussed in section~\ref{Ch6}~\cite{HLR,GHLR}.

\section{Eigenvalue-based reformulation}
\label{Ch3}
The direct numerical solution of the Richardson equations~\eqref{REq} becomes difficult when rapidities approach one another or a pair level. We therefore use the eigenvalue-based reformulation introduced in \cite{Faribault2011}. The essential idea is to replace the rapidities by the values of their resolvent at the pair energies.

\subsection{Eigenvalue-based equations}
In this section it is convenient to absorb the factor of two appearing in
the pair energies into the notation. Thus, we write
\begin{equation}
	\varepsilon_i \equiv 2\epsilon_i^{\rm(sp)},
	\qquad
	\lambda_j \equiv E_j,
\end{equation}
where $\epsilon_i^{\rm(sp)}$ are the single-particle energies used in
section~\ref{Ch2}. The Richardson equations then read
\begin{equation}
	\boxed{
		\frac{1}{g}
		-\sum_{i=1}^{L}\frac{1}{\varepsilon_i-\lambda_j}
		+2\sum_{\substack{k=1\\ k\neq j}}^{M}
		\frac{1}{\lambda_k-\lambda_j}=0,
		\qquad j=1,\ldots,M.
	}
	\label{eq:3_richardson}
\end{equation}
This is only a change of notation; the many-body energy remains
\begin{equation}
	E_{\mathcal U}=\sum_{j=1}^{M}\lambda_j.
	\label{eq:3_energy}
\end{equation}

We introduce the monic Baxter polynomial
\begin{equation}
	P(z)=\prod_{j=1}^{M}(z-\lambda_j),
	\label{Bax}
\end{equation}
and its logarithmic derivative
\begin{equation}
	\Lambda(z)=\frac{P'(z)}{P(z)}
	=\sum_{j=1}^{M}\frac{1}{z-\lambda_j}.
	\label{eq:3_Lambda_def}
\end{equation}
The double poles in $\Lambda'(z)+\Lambda^2(z)$ cancel, while its
simple pole at $z=\lambda_j$ has residue
\[
2\sum_{k\neq j}\frac{1}{\lambda_j-\lambda_k}.
\]
Using the Richardson equations, this residue can be expressed entirely
in terms of the pair energies. Equivalently, one obtains the
identity
\begin{equation}
	\Lambda'(z)+\Lambda^2(z)
	-2\sum_{j=1}^{M}
	\frac{F(\lambda_j)}{z-\lambda_j}=0,
	\label{eq:3_Riccati}
\end{equation}
where, for the Richardson model,
\begin{equation}
	F(\lambda_j)
	=-\frac12\sum_{i=1}^{L}\frac{1}{\varepsilon_i-\lambda_j}
	+\frac{1}{2g}.
	\label{eq:3_F}
\end{equation}
Equation~\eqref{eq:3_Riccati} is the Riccati form of the eigenvalue-based
representation. The numerical advantage is that the rapidities no longer
enter as unknowns in a system containing explicit rapidity--rapidity
singularities.

Taking the limit $z\to\varepsilon_i$ in~\eqref{eq:3_Riccati} and using
\eqref{eq:3_Lambda_def} gives a closed system for the $L$ quantities
$\Lambda(\varepsilon_i)$. After simplifying the terms with
$\varepsilon_i-\lambda_j$ and using
$\Lambda'(\varepsilon_i)=-\sum_j(\varepsilon_i-\lambda_j)^{-2}$, one finds
\begin{equation}
	\Lambda^2(\varepsilon_i)
	-\frac{1}{g}\Lambda(\varepsilon_i)
	+\sum_{k\neq i}^{L}
	\frac{\Lambda(\varepsilon_i)-\Lambda(\varepsilon_k)}
	{\varepsilon_k-\varepsilon_i}
	=0,
	\qquad i=1,\ldots,L.
	\label{eq:3_ebm_unscaled}
\end{equation}
Thus the original $M$ nonlinear Richardson equations are replaced by
$L$ coupled quadratic equations for the values of the resolvent on the
set of pair energies.

It is useful to introduce the dimensionless eigenvalue-based variables\footnote{
Since $g$ has the dimension of energy, whereas
$\Lambda(\varepsilon_i)$ has the dimension of inverse energy,
the rescaled variables $\Lambda_i$ are dimensionless.}
\begin{equation}
	\Lambda_i \equiv g\,\Lambda(\varepsilon_i).
	\label{eq:3_Lambda_scaled}
\end{equation}
Multiplication of~\eqref{eq:3_ebm_unscaled} by $g^2$ gives
\begin{equation}
	\boxed{
		\Lambda_i^2-\Lambda_i
		+g\sum_{k\neq i}^{L}
		\frac{\Lambda_i-\Lambda_k}{\varepsilon_k-\varepsilon_i}=0,
		\qquad i=1,\ldots,L.
	}
	\label{eq:3_ebm_quad}
\end{equation}
These equations are the nonlinear system solved by Newton--Raphson
iteration in the implementation described in section~\ref{Ch4}.

In the weak-coupling limit the rapidities approach occupied
pair levels. For a non-degenerate configuration, the
corresponding initial values can therefore be taken as
\begin{equation}
	\Lambda_i(0)=
	\begin{cases}
		1, & \varepsilon_i\ \text{occupied by one pair},\\[2pt]
		0, & \varepsilon_i\ \text{unoccupied},
	\end{cases}
	\label{eq:3_lambda_init}
\end{equation}
with
\begin{equation}
	\sum_{i=1}^{L}\Lambda_i(0)=M.
\end{equation}
For finite $g$ the solution is continued from a small nonzero coupling,
using the converged solution at one value of $g$ as the initial guess at
the next value. This continuation is the basis for the coupling scans
shown in section~\ref{Ch5}.

\subsection{Reconstruction of the rapidities}

Once the solution of~\eqref{eq:3_ebm_quad} is known, the rapidities are
reconstructed as the roots of the Baxter polynomial. We write the monic
polynomial in the standard form
\begin{equation}
	P(z)=z^M+a_1z^{M-1}+a_2z^{M-2}+\cdots+a_M
	= z^M+\sum_{m=1}^{M}a_m z^{M-m}.
	\label{eq:3_poly_expansion}
\end{equation}
Since $\Lambda(z)=P'(z)/P(z)$, at each pair energy we have
\begin{equation}
	\frac{\Lambda_i}{g}
	=\frac{P'(\varepsilon_i)}{P(\varepsilon_i)}.
	\label{eq:3_poly_ratio}
\end{equation}
Substituting~\eqref{eq:3_poly_expansion} into this relation gives the
linear equations
\begin{equation}
	\sum_{m=1}^{M-1}
	\left[
	\Lambda_i\varepsilon_i^{M-m}
	-g(M-m)\varepsilon_i^{M-m-1}
	\right]a_m
	+\Lambda_i a_M
	=
	gM\varepsilon_i^{M-1}-\Lambda_i\varepsilon_i^M,
	\qquad i=1,\ldots,L.
	\label{eq:3_poly_linear}
\end{equation}
Thus the reconstruction of $P(z)$ is a linear-algebra problem. The
system contains $L$ equations for the $M$ unknown coefficients
$a_1,\ldots,a_M$. In the current implementation, the first $M$ input
pair energies are used to form a square $M\times M$ system, which is
solved for the polynomial coefficients. The remaining $L-M$ equations
are not used in this reconstruction.

Finally, the rapidities are obtained from
\begin{equation}
	P(\lambda_j)=0,
	\qquad j=1,\ldots,M,
	\label{eq:3_root_reconstruction}
\end{equation}
and the many-body energy follows from~\eqref{eq:3_energy}. Root finding
is treated separately in section~\ref{Ch4}; in particular, the present
implementation uses the Laguerre method together with polynomial
deflation.

\section{Numerical implementation}
\label{Ch4}

The numerical procedure is designed to follow the solutions of the Richardson equations as the coupling $g$ is varied.  Its main ingredients are the eigenvalue-based equations~\eqref{eq:3_ebm_quad}, the linear reconstruction of the Baxter polynomial described in section~\ref{Ch3}, and a complex root finder based on Laguerre iteration with polynomial deflation.  The use of the eigenvalue-based variables avoids the direct treatment of the singular terms that appear when rapidities approach the pair levels or one another.

\subsection{Newton--Raphson iteration and linear algebra}

At a fixed value of $g$, the nonlinear system for the variables $\Lambda_i$ is written as
\begin{equation}
	\mathbf{F}(\boldsymbol{\Lambda})=\mathbf{0},
	\qquad
	F_i(\boldsymbol{\Lambda})
	=\Lambda_i^2-\Lambda_i
	+g\sum_{k\neq i}
	\frac{\Lambda_i-\Lambda_k}{\varepsilon_k-\varepsilon_i} .
	\label{eq:4_num-F}
\end{equation}
The Newton--Raphson step is obtained from
\begin{equation}
	\mathbf{J}(\boldsymbol{\Lambda})\,\delta\boldsymbol{\Lambda}
	=-\mathbf{F}(\boldsymbol{\Lambda}),
	\qquad
	J_{ij}=\frac{\partial F_i}{\partial\Lambda_j},
	\label{eq:4_num-Newton}
\end{equation}
and the solution is updated according to
\begin{equation}
	\boldsymbol{\Lambda}\longleftarrow
	\boldsymbol{\Lambda}+\delta\boldsymbol{\Lambda}.
\end{equation}
In the current C++ implementation, the Jacobian entries are evaluated by forward finite differences,
\begin{equation}
	J_{ij}\simeq\frac{F_i(\boldsymbol{\Lambda}+h\mathbf e_j)-F_i(\boldsymbol{\Lambda})}{h},
	\qquad h=10^{-6},
	\label{eq:4_num-Jacobian}
\end{equation}
where $\mathbf e_j$ is the $j$th unit vector.  Each Newton step therefore reduces to the solution of a linear system.  In the 
present implementation this system is solved by LU decomposition.  Writing
\begin{equation}
	\mathbf{J}=\mathbf{L}\mathbf{U},
	\qquad
	\mathbf{L}\mathbf{y}=-\mathbf{F},
	\qquad
	\mathbf{U}\,\delta\boldsymbol{\Lambda}=\mathbf{y},
\end{equation}
the triangular systems are solved by forward and backward substitution.

The current implementation uses an update-based stopping test: iteration terminates when
$\sqrt{\left|\sum_i(\delta\Lambda_i)^2\right|}\leq10^{-5}$.  No separate residual-norm test is imposed in the present code.  For 
the continuation in $g$, the converged solution at the previous coupling is used as the initial guess at the next one.  At $g=0$ 
the initial values are taken from~\eqref{eq:3_lambda_init}; the first nonzero coupling is then used to start the continuation.  
This procedure is particularly important near configurations where a direct solution of the Richardson equations becomes 
numerically ill-conditioned.

\subsection{Polynomial reconstruction and root finding}
Once the values $\Lambda(\varepsilon_i)$ have been obtained, the coefficients of the Baxter polynomial are determined from the 
linear system~\eqref{eq:3_poly_linear}.  The current implementation selects the first $M$ input pair energies and solves the 
resulting square $M\times M$ system for the $M$ polynomial coefficients.  It does not perform a least-squares fit to all $L$ 
equations.  The polynomial is then reconstructed as (\ref{Bax})
with the normalization fixed by its leading coefficient.  The rapidities are recovered as the roots of $P(z)$.

Because the rapidities may be real or occur in complex-conjugate pairs, the implementation uses a complex root-finding procedure.  
For a polynomial $P_n(x)$ of degree $n$, define
\begin{equation}
	G(x)=\frac{P_n'(x)}{P_n(x)},
	\qquad
	H(x)=G^2(x)-\frac{P_n''(x)}{P_n(x)}.
\end{equation}
In the implementation, the first and second derivatives entering $G$ and $H$ are themselves evaluated by forward finite 
differences, with step $h_z=10^{-5}$.  The Laguerre update can then be written as~\cite{NumericalRecipes}
\begin{equation}
	\delta x
	=
	\frac{n}{G(x)\pm
		\sqrt{(n-1)\left[nH(x)-G^2(x)\right]}},
	\qquad
	x\longleftarrow x-\delta x,
	\label{eq:4_num-Laguerre}
\end{equation}
where the sign is selected so that the denominator has the larger magnitude.  The root search is initialized at $x=5$ for each 
deflation stage and the code terminates an iteration when $|\delta x|\leq10^{-5}$.  The present implementation uses the update 
size as its stopping test and does not impose a separate polynomial-residual threshold.

After one root $r$ has been found, polynomial deflation is used to construct
\begin{equation}
	P_n(x)=(x-r)P_{n-1}(x).
\end{equation}
The Laguerre procedure is then applied to the reduced polynomial, and the process is repeated until all $M$ rapidities have been 
obtained.  This procedure avoids having to solve for all roots simultaneously and naturally accommodates complex roots.

\subsection{Algorithmic workflow}
For a given single-particle spectrum and number of pairs, the complete calculation proceeds as follows.  The input data specify 
the distinct pair energies, the number of pairs, and the range of coupling constants to be explored.  Starting from the $g=0$ 
configuration, the program solves~\eqref{eq:4_num-F} by Newton--Raphson iteration.  The resulting $\Lambda_i$ are inserted into 
the linear polynomial reconstruction equations~\eqref{eq:3_poly_linear}.  The roots of the reconstructed polynomial are then 
obtained by Laguerre iteration and deflation.  The resulting rapidities determine the many-body energy through
(\ref{eq:3_energy}) for the paired sector considered here.

The next value of $g$ is reached by incrementing the coupling and using the previous converged $\boldsymbol{\Lambda}$ as the 
initial condition.  Thus the numerical calculation follows a continuous branch of the exact Richardson solution rather than 
solving each coupling value independently.  The present workflow is restricted to non-degenerate pair energies.

\subsection{Code structure and numerical performance}
\label{code}
The implementation is written in C++ and consists of the files \texttt{main.cpp}, \texttt{solver.cpp}, and \texttt{solver.h}.  
The numerical functionality is organized into three principal classes: \texttt{cSqMatrix} for linear systems, \texttt{cEquations} 
for the nonlinear eigenvalue-based equations, and \texttt{cPolynomial} for polynomial construction and root finding.  The source 
code and instructions for preparing the input files are publicly available at
\url{https://github.com/GrzegorzBiskowski/Richardson-solver}.
The program outputs the computed energies as functions of $g$ in both text and CSV formats.

The code uses complex arithmetic throughout, which allows the same implementation to treat real rapidities and complex conjugate 
pairs without changing the numerical workflow.  The present implementation was tested for systems containing up to 20 Cooper 
pairs.  For larger systems, numerical discontinuities can appear in the $\Lambda_i$ continuation, particularly near 
ill-conditioned configurations.  The algorithms are intentionally kept simple, so that such limitations are transparent and the 
individual numerical components can be replaced by higher precision or more sophisticated routines in future developments.

\section{Numerical examples}
\label{Ch5}
We first illustrate the numerical procedure for several representative non-degenerate spectra.  The purpose of these examples is 
twofold: to test the continuation of the eigenvalue-based variables and the reconstruction of the rapidities, and to illustrate 
how the distribution of non-degenerate pair energies affects the trajectories of the Richardson roots.  Unless stated otherwise, 
we consider the paired sector with $M$ Cooper pairs and use the convention of section~\ref{Ch3}, in which $\varepsilon_j$ denotes 
the corresponding pair energy at $g=0$.

\subsection{Non-degenerate spectra}
\begin{figure}[h]
	\centering
	\begin{minipage}{.45\textwidth}
		\centering
		\scalebox{0.45}{\input{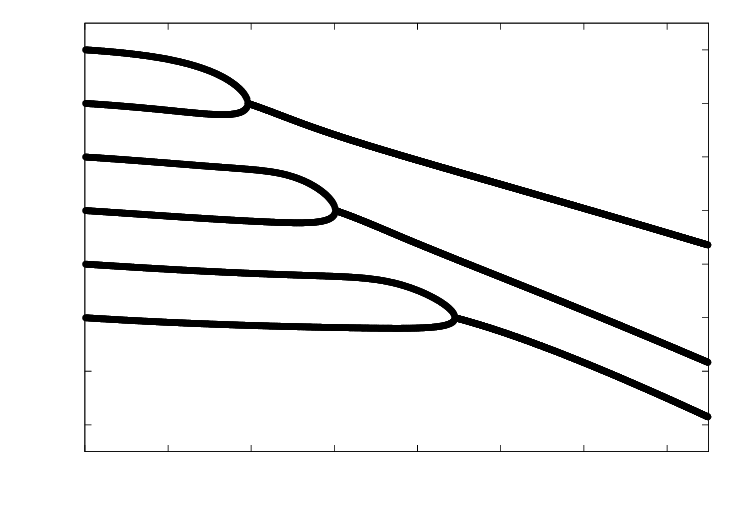}}
		\caption{Real parts of the Richardson rapidities for the picket-fence spectrum with $L=12$ pair levels and $M=6$ Cooper pairs.}
		\label{fig:picket_fence_re}
	\end{minipage}\text{\qquad}%
	\begin{minipage}{.45\textwidth}
		\centering
		\scalebox{0.45}{\input{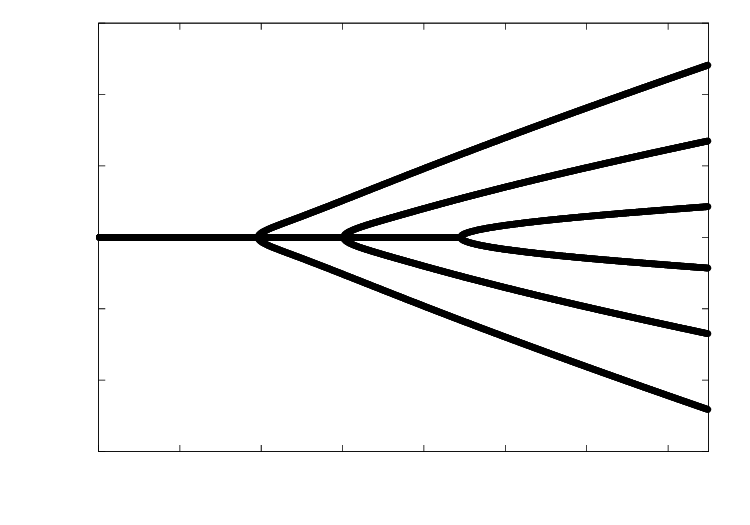}}
		\caption{Imaginary parts of the Richardson rapidities for the picket-fence spectrum with $L=12$ pair levels and $M=6$ Cooper pairs.}
		\label{fig:picket_fence_im}
	\end{minipage}
\end{figure}

As a reference case, we consider the equally spaced spectrum
\begin{equation}
	\varepsilon_j=j+\tfrac12,
	\qquad j=0,1,\ldots,11,
\end{equation}
with $M=6$ pairs occupying the lowest six levels.  We refer to this as the picket-fence spectrum.  The real and imaginary parts 
of the resulting rapidities are shown in figures~\ref{fig:picket_fence_re} and~\ref{fig:picket_fence_im}.

For weak coupling the rapidities are close to the occupied pair levels.  As $g$ increases, neighbouring real rapidities can 
approach one another and subsequently leave the real axis as complex conjugate pairs.  This behaviour is a direct manifestation 
of the singular structure of the Richardson equations and is precisely the regime in which the eigenvalue-based continuation is 
useful.  Because the Hamiltonian is Hermitian, the rapidities occur in configurations whose sum, and hence the total many-body 
energy, remains real.

\begin{figure}[h]
	\centering
	\begin{minipage}{.45\textwidth}
		\centering
		\scalebox{0.45}{\input{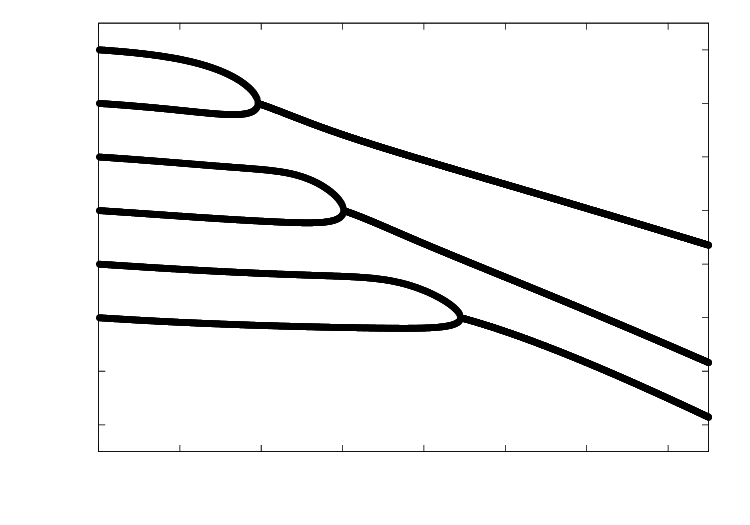}}
		\caption{Real parts of the rapidities for the spectrum $\varepsilon_j=10(j+\tfrac12)$ with the coupling rescaled by the same factor.}
		\label{fig:10g_picket_fence_re}
	\end{minipage}\text{\qquad}%
	\begin{minipage}{.45\textwidth}
		\centering
		\scalebox{0.45}{\input{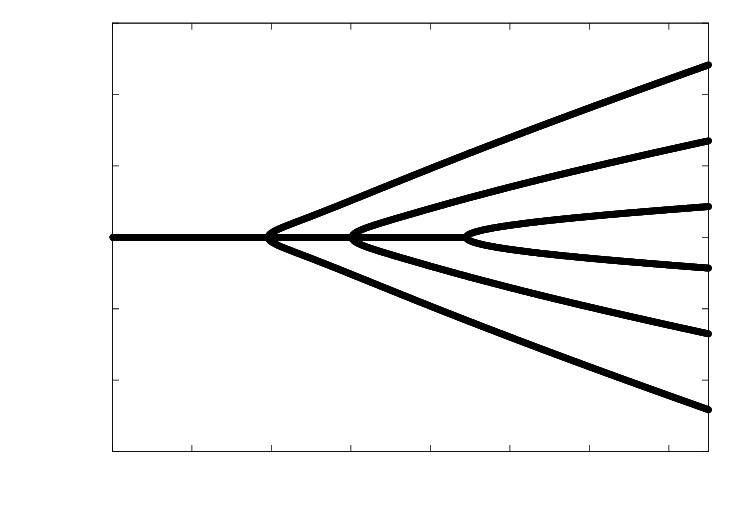}}
		\caption{Imaginary parts of the rapidities for the rescaled picket-fence spectrum.}
		\label{fig:10g_picket_fence_im}
	\end{minipage}
\end{figure}

The scale invariance of the Richardson equations provides a simple numerical check.  Under a simultaneous rescaling
\begin{equation}
	\varepsilon_j\longrightarrow s\varepsilon_j,
	\qquad
	g\longrightarrow sg,
\end{equation}
all rapidities and many-body energies scale by the same factor $s$.  Thus the dimensionless trajectories are unchanged.  As an 
explicit check, we rescale the equally spaced spectrum by a factor of ten and use the corresponding rescaled coupling.  The 
resulting rapidity trajectories coincide with those of the reference calculation after the overall energy rescaling.\footnote{
All rapidities and many-body energies scale by the same factor $s$. Consequently, the rapidity trajectories coincide after the 
energy and coupling axes are rescaled accordingly.} 
The corresponding figures are shown in figures~\ref{fig:10g_picket_fence_re} and~\ref{fig:10g_picket_fence_im}.

\begin{figure}[h]
	\centering
	\begin{minipage}{.45\textwidth}
		\centering
		\scalebox{0.45}{\input{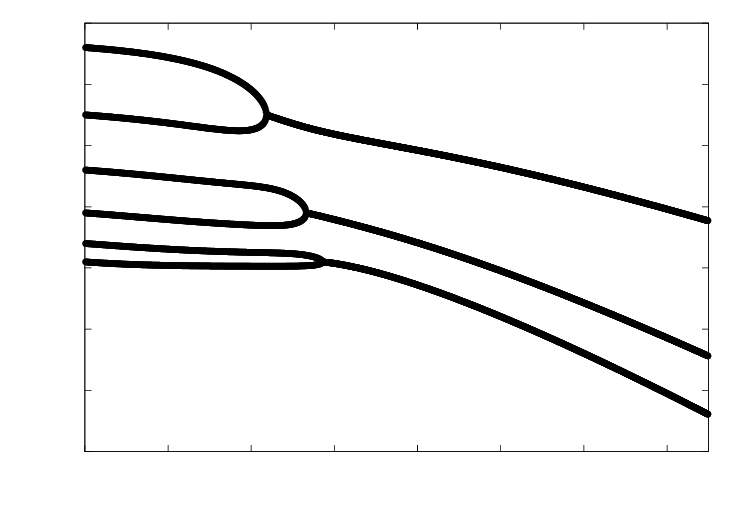}}
		\caption{Real parts of the rapidities for the one-dimensional infinite square well spectrum with $L=12$ and $M=6$.}
		\label{fig:square_well_re}
	\end{minipage}\text{\qquad}%
	\begin{minipage}{.45\textwidth}
		\centering
		\scalebox{0.45}{\input{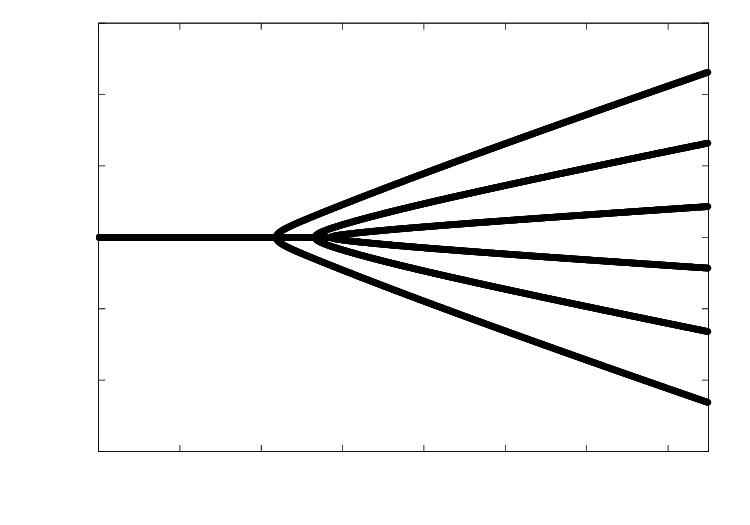}}
		\caption{Imaginary parts of the rapidities for the one-dimensional infinite square well spectrum with $L=12$ and $M=6$.}
		\label{fig:square_well_im}
	\end{minipage}
\end{figure}

We next consider the spectrum of a one-dimensional infinite square well, for which the pair energies are proportional to $j^2$,
\begin{equation}
	\varepsilon_j\propto j^2,
	\qquad j=1,2,\ldots,12.
\end{equation}
The resulting real and imaginary parts of the rapidities are shown in figures~\ref{fig:square_well_re} 
and~\ref{fig:square_well_im}.  
In contrast with the picket-fence spectrum, the level spacings increase with $j$. Accordingly, the numerical continuation
reveals that neighbouring rapidities approach one another at coupling values that are not equally spaced. This example
illustrates that the continuation algorithm does not rely on an equally spaced single-particle spectrum.

\begin{figure}[h]
    \centering
    \begin{minipage}{.45\textwidth}
        \centering
        \scalebox{0.45}{\input{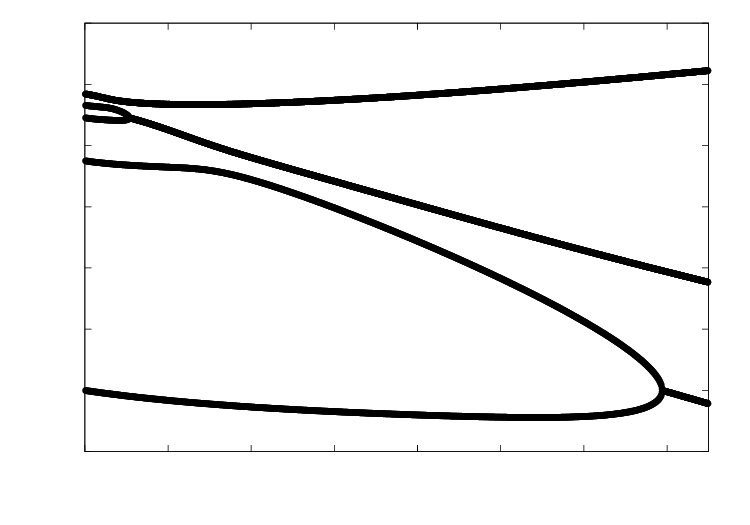}}
        \caption{Real parts of the Richardson rapidities for
        the hydrogen-like pair-energy spectrum
        $\varepsilon_j=-\varepsilon_{\mathrm H}j^{-2}$,
        with $L=M=6$.}
        \label{fig:inv_square_re}
    \end{minipage}\text{\qquad}%
    \begin{minipage}{.45\textwidth}
        \centering
        \scalebox{0.45}{\input{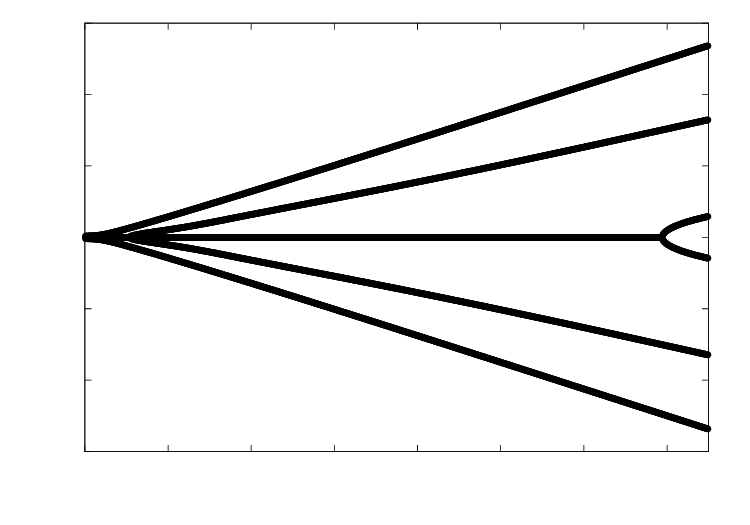}}
        \caption{Imaginary parts of the Richardson rapidities
        for the hydrogen-like pair-energy spectrum
        $\varepsilon_j=-\varepsilon_{\mathrm H}j^{-2}$,
        with $L=M=6$.}
        \label{fig:inv_square_im}
    \end{minipage}
\end{figure}

Finally, we consider a hydrogen-like pair-energy spectrum,
\begin{equation}
    \varepsilon_j=-\frac{\varepsilon_{\mathrm H}}{j^2},
    \qquad j=1,2,\ldots,6,
\end{equation}
where $\varepsilon_{\mathrm H}>0$ sets the energy scale.
In the numerical calculations presented below, we set
$\varepsilon_{\mathrm H}=5$. We take $M=L=6$, so that all
available pair levels are occupied in the non-interacting
reference configuration. Consequently, the seniority-zero
sector contains only a single many-body state,
$\mathcal{D}_M=\binom{6}{6}=1$. The corresponding rapidity
trajectories are shown in figures~\ref{fig:inv_square_re}
and~\ref{fig:inv_square_im}. This example is therefore
intended primarily to illustrate the evolution of the
Richardson rapidities and the reconstruction of the
corresponding roots for a strongly non-uniform spectrum.

For the fully occupied configuration, the exact many-body
energy follows directly from the Hamiltonian~\eqref{HU}:
\begin{equation}
    E_{\mathrm{full}}(g)
    =\sum_{j=1}^{6}\varepsilon_j-6g
    =-\frac{5369}{3600}\varepsilon_{\mathrm H}-6g.
\end{equation}
This provides an independent check of the numerical
reconstruction, since the sum of the six rapidities must
satisfy
\begin{equation}
    \sum_{\alpha=1}^{6}\lambda_\alpha(g)
    =E_{\mathrm{full}}(g).
\end{equation}
We have verified this identity numerically throughout the
coupling continuation. At each value of $g$, the sum of the
six reconstructed rapidities reproduces the analytical
linear dependence of $E_{\mathrm{full}}(g)$ on the coupling,
with agreement to numerical precision. This confirms the
consistency of the rapidity reconstruction for the fully
occupied configuration.

The adjacent pair-energy spacings,
\begin{equation}
    \varepsilon_{j+1}-\varepsilon_j
    =\varepsilon_{\mathrm H}
    \frac{2j+1}{j^2(j+1)^2},
\end{equation}
decrease with increasing $j$. This hierarchy is reflected
in the rapidity trajectories: rapidities originating from
neighbouring high-lying pair levels approach one another
at smaller coupling strengths than those associated with
the more widely separated low-lying levels. The example
thus illustrates how the detailed geometry of the
single-particle spectrum influences the coupling dependence
of the Richardson roots.

\subsection{Exact finite-temperature thermodynamics}
\label{sec:finiteT}
The discrete spectrum obtained from the Richardson equations provides a direct route to the thermodynamics of finite systems.  In 
contrast to a thermodynamic limit treatment, no mean-field critical temperature or continuum approximation is required.  
For a finite system there is no nonanalytic phase transition; instead, thermodynamic crossovers are encoded in the
temperature dependence of observables constructed from the exact discrete spectrum.

We restrict the discussion to the even-particle, seniority-zero sector with $N=2M$ particles.  Let
\[
\mathsf E_M(g)=
\{E_\alpha(g)\}_{\alpha=0}^{\mathcal D_M-1}.
\]
denote the complete set of many-body eigenenergies in this sector.  For $L$ non-degenerate pair levels, the dimension of this 
sector is
\begin{equation}
	\mathcal{D}_M=\binom{L}{M}.
\end{equation}
The finite-temperature calculation must include all these many-body eigenstates at fixed particle number, rather than only the 
ground-state Richardson solution.

The complete spectrum can be organized naturally within the eigenvalue-based formulation.  At $g=0$, every occupation pattern of 
$M$ pair levels defines an eigenstate, suggesting a natural continuation strategy in which each occupation pattern is followed 
separately.  The corresponding initial values of the eigenvalue-based variables are therefore
\begin{equation}
	\Lambda_i(g=0)=
	\begin{cases}
		1, & \varepsilon_i\ \text{occupied},\\[2pt]
		0, & \varepsilon_i\ \text{unoccupied},
	\end{cases}
\end{equation}
for each allowed occupation pattern.\footnote{The values at $g=0$ are understood as the continuous
$g\to 0$ limits of the rescaled eigenvalue-based variables.} 
Continuing each such solution in $g$ generates the interacting spectrum.  In the finite 
systems considered below, the complete spectra are checked independently by exact diagonalization of the paired Hamiltonian.

The canonical partition function in the seniority-zero sector is then
\begin{equation}
	Z_M(\beta,g)=
	\sum_{\alpha=0}^{\mathcal{D}_M-1}
	\exp[-\beta E_\alpha(g)],
	\qquad
	\beta=\frac{1}{k_{\rm B}T},
	\label{eq:5_Z}
\end{equation}
where the index $\alpha$ labels individual many-body eigenstates; hence any degeneracies are already included through repeated 
eigenvalues in the state list.  In the calculations presented below we use units with $k_{\rm B}=1$.  The Helmholtz free energy, 
internal energy, entropy, and heat capacity follow from
\begin{align}
	F_M(T,g)&=-T\ln Z_M,
	\label{eq:5_F}
	\\
	U_M(T,g)&=-\frac{\partial}{\partial\beta}\ln Z_M
	=\frac{1}{Z_M}\sum_\alpha E_\alpha\,{\rm e}^{-\beta E_\alpha},
	\label{eq:5_U}
	\\
	S_M(T,g)&=\frac{U_M-F_M}{T},
	\label{eq:5_S}
	\\
	C_M(T,g)&=\left.\frac{\partial U_M}{\partial T}\right|_M
	=\frac{\langle E^2\rangle-\langle E\rangle^2}{T^2}.
	\label{eq:5_C}
\end{align}
Thus the finite-size heat capacity is determined directly by fluctuations of the exact many-body energy.

A second useful observable is the pairing contribution to the internal energy,
\begin{equation}
	E_p(T,g)=U_M(T,g)-U_M(T,0),
	\label{eq:5_Ep}
\end{equation}
which reduces to the conventional pairing-energy definition used in studies of finite Richardson 
systems~\cite{Ciechan}.\footnote{The reference energy $U_M(T,0)$ is evaluated at the
same temperature and fixed particle number. Thus, $E_p$
measures the interaction-induced change in the total internal
energy and should not be identified with the expectation
value of the interaction term alone.}
Its temperature derivative obeys the exact identity
\begin{equation}
	\frac{\partial E_p(T,g)}{\partial T}
	=C_M(T,g)-C_M(T,0).
	\label{eq:5_Ep_derivative}
\end{equation}
Consequently, the thermal evolution of the pairing energy and the modification of the heat capacity by the interaction are not 
independent observables.

The coupling derivative of the free energy gives a direct measure of pair correlations.  For the Hamiltonian~\eqref{HU}, define
\begin{equation}
	\mathcal{P}_M(T,g)
	\equiv-\frac{\partial F_M}{\partial g}.
	\label{eq:5_paircorr}
\end{equation}
By the Hellmann--Feynman theorem,
\begin{equation}
	\mathcal{P}_M(T,g)
	=\left\langle B_0^\dagger B_0\right\rangle_T,
	\qquad
	B_0^\dagger=\sum_j b_j^\dagger,
	\label{eq:5_paircorr_HF}
\end{equation}
so that $\mathcal{P}_M$ probes the thermal expectation value of the pair-transfer operator.  This quantity complements the 
pairing energy because it is directly tied to the interaction term of the Hamiltonian rather than only to the difference of two 
internal energies.\footnote{Since $B_0^\dagger B_0=\sum_{j,k}b_j^\dagger b_k$, this operator contains both diagonal 
pair-occupation terms and off-diagonal pair-transfer correlations. In the fixed-$M$ seniority-zero sector, the diagonal 
contribution equals $M$, while the off-diagonal terms encode correlations between distinct pair levels.}

For a general even-particle canonical calculation, the same construction can be extended beyond the seniority-zero sector by 
summing over blocked configurations.  If $\mathcal{B}$ denotes a set of blocked levels and $M=(N-|\mathcal{B}|)/2$, the 
contribution of that sector has the form
\begin{equation}
	Z_{N,\mathcal{B}}
	=d_{\mathcal{B}}\,{\rm e}^{-\beta E_{\mathcal{B}}}
	Z_{M,\mathcal{B}}^{\rm pair},
	\label{eq:5_blockedZ}
\end{equation}
with $E_{\mathcal B}=\sum_{i\in\mathcal B}\varepsilon_i$
and $d_{\mathcal B}$ accounting for the spin multiplicity
of the blocked states.\footnote{For spin-degenerate time-reversed single-particle
levels and in the absence of a magnetic field, each blocked
level contains one unpaired fermion whose spin can occupy
either of the two time-reversed states. If the canonical
trace includes all spin configurations, the corresponding
multiplicity is therefore
$d_{\mathcal B}=2^{|\mathcal B|}$. If a specific spin
configuration or spin projection is fixed, the multiplicity
must be adjusted accordingly.}
The full fixed-$N$ partition function is obtained by summing 
these contributions over all allowed blocked sectors.  The calculations below use the seniority-zero sector, for which the 
spectrum can be enumerated transparently and the role of the single-particle spectral geometry can be isolated.

\subsection{Spectral geometry and thermal anomalies}
\label{sec:thermal_geometry}
The finite system is characterized not by a single critical temperature but by a hierarchy of discrete many-body excitation 
scales.  We denote the excitation gaps above the ground state by
\begin{equation}
	\Delta_\alpha(g)=E_\alpha(g)-E_0(g),
	\qquad \alpha=1,\ldots,\mathcal{D}_M-1.
	\label{eq:5_gaps}
\end{equation}
The positions of extrema in $C_M(T,g)$ therefore depend on the distribution of the low-lying gaps.  Since those gaps are 
inherited from the underlying pair-energy spectrum, changing the spectral geometry changes the thermal response even when the 
interaction Hamiltonian is otherwise identical.

This observation suggests a useful distinction between two types of finite-size thermal scales.  A low-temperature anomaly may be 
associated with the smallest many-body gaps and the rearrangement of strongly correlated low-lying states, whereas a second, 
higher-temperature anomaly can reflect a broader set of discrete excitations associated with larger single-particle and many-body 
spacings.  We stress that neither feature is a phase transition in a finite system.

\begin{figure}[htbp]
	\centering
	\scalebox{0.78}{\input{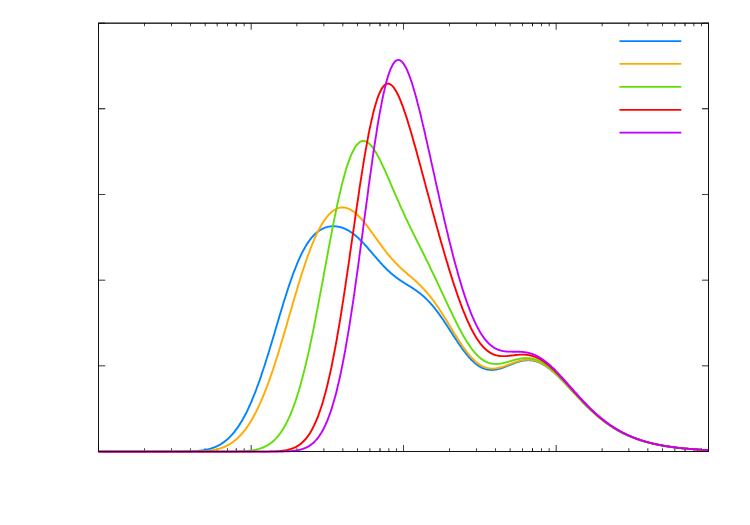}}
	\caption{Heat capacity $C_M(T,g)$ for the finite Richardson model with $L=8$, $M=4$ and the hydrogen-like spectrum~\eqref{eq:5_hydrogen}.  For weak coupling the heat capacity develops two well-separated maxima.  The two anomalies move towards each other as the pairing interaction is increased.}
	\label{fig:finiteT_hydrogen_Cv}
\end{figure}

As a first systematic test, we consider a hydrogen-like single-particle spectrum with $L=8$ levels and $M=4$ pairs,
\begin{equation}
	\epsilon_j^{\rm(sp)}=-\frac{1}{(j+1)^2},
	\qquad j=0,1,\ldots,7,
	\qquad \varepsilon_j=2\epsilon_j^{\rm(sp)}=-\frac{2}{(j+1)^2},
	\label{eq:5_hydrogen}
\end{equation}
where $\varepsilon_j$ is the pair energy used in the Richardson equations. At $g=0$, the first excitation replaces the pair in level $j=3$ by one in level $j=4$, giving $\Delta_1=2(1/16-1/25)=0.045$, as listed in Table~\ref{tab:finiteT_hydrogen}.
The seniority-zero Hilbert space contains $\binom{8}{4}=70$ many-body states.  The finite-temperature quantities are obtained from the complete discrete spectrum.  Figure~\ref{fig:finiteT_hydrogen_Cv} shows the resulting heat capacity for several weak and intermediate couplings.

At $g=0$ the two maxima occur at approximately
\begin{equation}
	T_1\simeq0.035,
	\qquad
	T_2\simeq0.656,
\end{equation}
while for $g=0.02$ they are shifted to approximately
\begin{equation}
	T_1\simeq0.054,
	\qquad
	T_2\simeq0.638.
\end{equation}

\begin{table}[htbp]
	\centering
	\begin{tabular}{c c c c c}
		\hline
		$g$ & $\Delta_1$ & $T_1$ & $C_M(T_1)$ & $T_2$ \\
		\hline
		$0$     & $0.0450$ & $0.0346$ & $1.315$ & $0.6563$ \\
		$0.010$ & $0.0513$ & $0.0398$ & $1.424$ & $0.6524$ \\
		$0.020$ & $0.0963$ & $0.0543$ & $1.812$ & $0.6382$ \\
		$0.030$ & $0.1652$ & $0.0790$ & $2.146$ & $0.6016$ \\
		$0.035$ & $0.2029$ & $0.0924$ & $2.285$ & $0.5401$ \\
		\hline
	\end{tabular}
	\caption{Low-temperature gap $\Delta_1=E_1-E_0$ and the two heat-capacity anomalies for the hydrogen-like $L=8$, $M=4$ system.}
	\label{tab:finiteT_hydrogen}
\end{table}
As the pairing strength $g$ increases, the low-temperature peak becomes more pronounced and shifts towards higher temperatures, whereas the high-temperature peak progressively flattens. At $g=0.035$, two distinct local maxima are still present, located at approximately $T_1\simeq 0.092$ and $T_2\simeq 0.540$. However, at $g\simeq 0.03515$, the high-temperature peak is no longer present as a local maximum. Thus, the disappearance of the secondary peak occurs within the interval $0.035<g_c<0.03515$, rather than through the merging of the two maxima.

\begin{figure}[htbp]
	\centering
	\scalebox{0.78}{\input{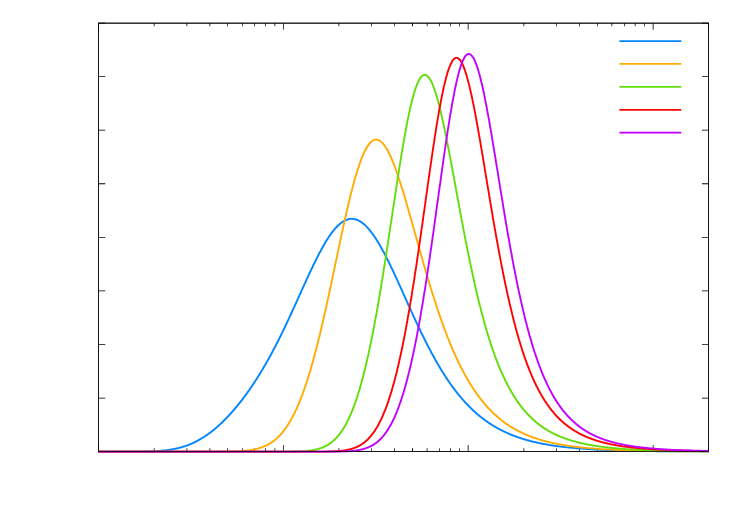}}
	\caption{Heat capacity $C_M(T,g)$ for the finite Richardson model with $L=8$, $M=4$ and the picket-fence spectrum $\varepsilon_j = 2j$.}
	\label{fig:finiteT_picketFence_Cv}
\end{figure}

\begin{figure}[htbp]
	\centering
	\scalebox{0.78}{\input{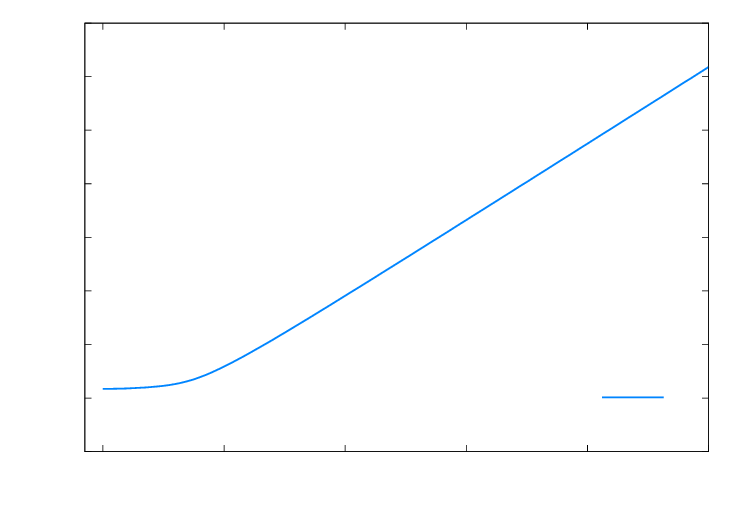}}
	\caption{Temperatures of the heat-capacity anomalies for the picket-fence $L=8$, $M=4$ system.}
	\label{fig:finiteT_picketFence_peaks}
\end{figure}

\begin{figure}[htbp]
	\centering
	\scalebox{0.78}{\input{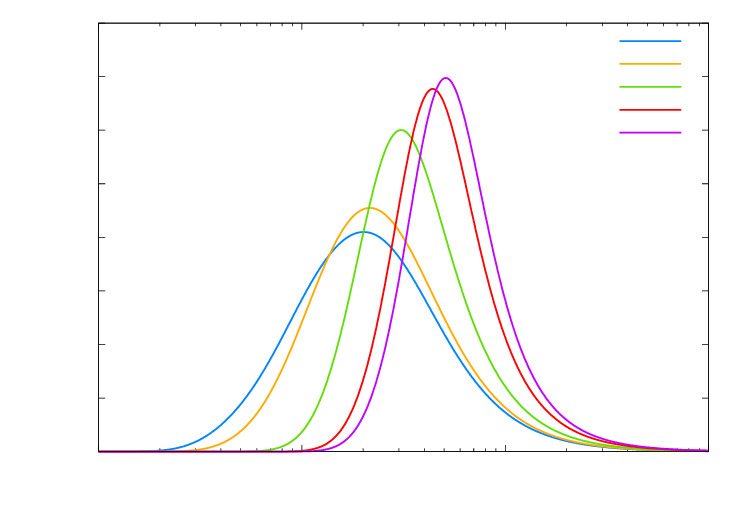}}
	\caption{Heat capacity $C_M(T,g)$ for the finite Richardson model with $L=8$, $M=4$ and the quadratic $\varepsilon_j = j^2$ spectrum.}
	\label{fig:finiteT_quadratic_Cv}
\end{figure}

\begin{figure}[htbp]
	\centering
	\scalebox{0.78}{\input{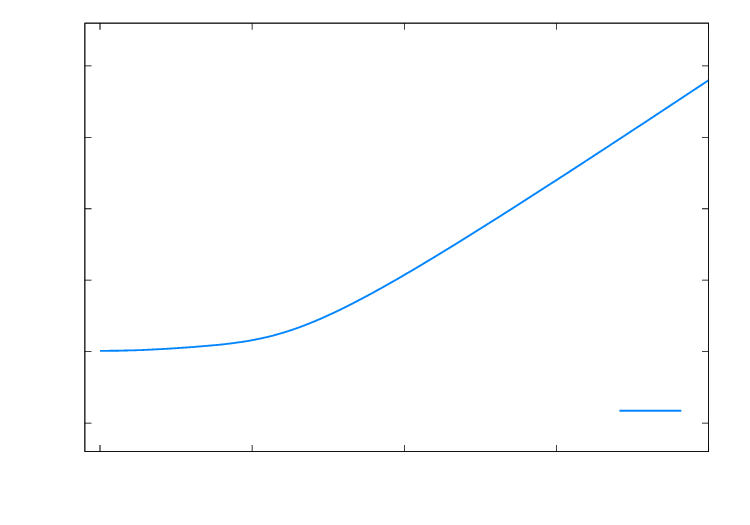}}
	\caption{Temperatures of the heat-capacity anomalies for the quadratic $L=8$, $M=4$ system.}
	\label{fig:finiteT_quadratic_peaks}
\end{figure}

The effect is not a generic consequence of finite size alone. For comparison, the equally spaced picket-fence spectrum with $L=8$ and $M=4$ has a single dominant heat-capacity maximum throughout the coupling interval displayed in Figs.~\ref{fig:finiteT_picketFence_Cv} and~\ref{fig:finiteT_picketFence_peaks}. At $g=2.0$ its position is approximately $T\simeq5.81$. The hydrogen-like and picket-fence examples have different pair-energy scales, so this comparison concerns the qualitative structure of the thermal response rather than equal numerical values of $g$ or $T$. The quadratic spectrum also exhibits a single dominant maximum over the range displayed in Figs.~\ref{fig:finiteT_quadratic_Cv} and~\ref{fig:finiteT_quadratic_peaks}.

\begin{figure}[htbp]
	\centering
	
	\scalebox{0.78}{\input{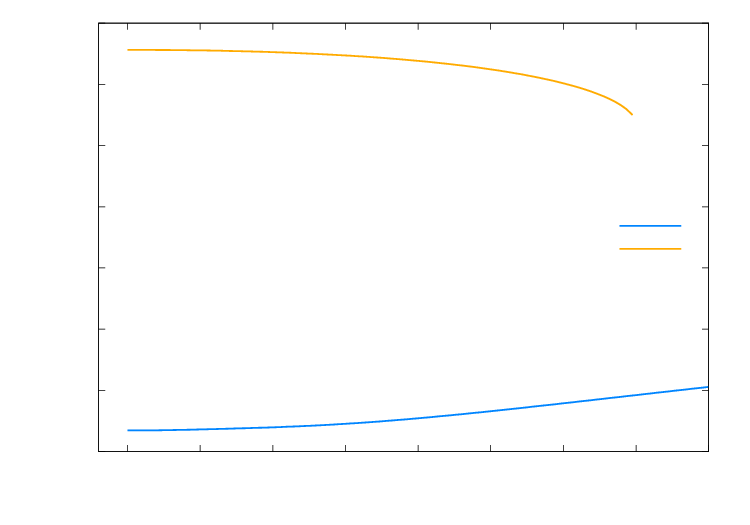}}
	\caption{Temperatures of the low- and high-temperature heat-capacity anomalies for the hydrogen-like $L=8$, $M=4$ system.  The higher-temperature branch terminates when the corresponding local maximum ceases to exist; the two branches do not merge.}
	\label{fig:finiteT_hydrogen_peaks}
\end{figure}

The evolution of the two anomaly temperatures provides a compact characterization of the crossover.  We denote the two local maxima by $T_1(g)<T_2(g)$ whenever both are resolved.  Figure~\ref{fig:finiteT_hydrogen_peaks} shows their trajectories.  The anomaly splitting,
\begin{equation}
	\mathcal{R}_T(g)=\frac{T_2(g)}{T_1(g)},
	\label{eq:5_peak_ratio}
\end{equation}
decreases rapidly as the coupling is increased.  In this sense the pairing interaction continuously reorganizes the thermal spectrum until the higher-temperature local maximum ceases to exist as a separate extremum.

The preliminary results already indicate a useful physical interpretation.  The first anomaly tracks the rapidly increasing 
lowest many-body gap generated by the pairing interaction, as illustrated by the simultaneous evolution of $\Delta_1$ and $T_1$ 
in Table~\ref{tab:finiteT_hydrogen}.  The second anomaly moves much more slowly and therefore probes a larger excitation scale of 
the discrete spectrum.  The separation of the two scales is particularly pronounced for the hydrogen-like spectrum because its 
pair-energy spacings are strongly non-uniform.  This provides a concrete example in which the geometry of the finite pair-energy 
spectrum controls the structure of the thermodynamic response.

It is also important that a double maximum in $C_M(T,g)$
does not by itself imply re-entrant pairing. The heat
capacity probes thermal fluctuations of the many-body
energy, whereas the pair-transfer observable
$\mathcal{P}_M(T,g)$ provides a direct measure of pairing
correlations. Consequently, the occurrence of multiple
heat-capacity anomalies does not establish a thermal
enhancement or re-entrance of pairing correlations.
A systematic investigation of the temperature dependence
of $\mathcal{P}_M(T,g)$, particularly in partially filled
or more strongly shell-separated spectra, is left as a
natural extension of the present study.

The finite-temperature construction therefore provides a natural extension of the numerical framework from individual Richardson 
roots to complete finite spectra. Once the spectrum is enumerated and inserted into the canonical partition function, the same 
framework can be used to investigate how non-uniform level spacing, particle number, filling, and interaction strength jointly 
determine finite-size thermodynamic anomalies. The hydrogen-like example is intended as a first demonstration of this programme 
rather than as an exhaustive classification of the possible behaviours.
	
\section{Discussion}
\label{Ch6}

We have presented a numerical workflow for the reduced BCS Richardson model that combines the eigenvalue-based formulation of the Bethe equations, reconstruction of the Baxter polynomial, and complex root finding. The method follows solutions as the pairing strength varies and avoids direct iteration of the Richardson equations in their singular rapidity form. Its contribution is a transparent combination of established equations and standard numerical procedures, together with spectral and thermodynamic benchmarks for non-degenerate single-particle spectra.

The finite Richardson spectrum also provides direct access to canonical thermodynamic quantities at fixed particle number. For the hydrogen-like system with $L=8$ levels and $M=4$ pairs, the heat capacity has two resolved maxima at weak coupling. The lower-temperature maximum shifts from approximately $T_1=0.0346$ at $g=0$ to $T_1=0.0924$ at $g=0.035$, while the higher-temperature maximum moves from $T_2=0.6563$ to $T_2=0.5401$. At $g\simeq0.03515$ the higher-temperature maximum is no longer present as a distinct local extremum at the numerical resolution used. The equally spaced and quadratic comparison spectra, over the parameter ranges displayed, exhibit a single dominant heat-capacity maximum. These results illustrate how the distribution of single-particle level spacings can influence the thermal response of a finite pairing system. The observed features are smooth finite-size crossovers, not thermodynamic phase transitions.

The heat-capacity structure should not by itself be interpreted
as evidence for re-entrant pairing. The heat capacity probes
thermal fluctuations of the many-body energy, whereas the
temperature dependence of pairing correlations requires
a separate analysis of an appropriate correlation observable,
such as the pair-transfer expectation value
$\mathcal{P}_M(T,g)$. The double maxima found in the present
calculations therefore establish multiple thermal anomalies
in the finite-size energy fluctuations, but do not establish
a corresponding enhancement or re-entrance of pairing
correlations. A systematic investigation of the temperature
dependence of $\mathcal{P}_M(T,g)$, including its possible
non-monotonic behavior in different spectral geometries,
is left for future work. More generally, the results presented
here do not imply that double heat-capacity maxima are generic
features of Richardson--Gaudin models.

Several limitations should be kept in view. The implementation has been tested for systems containing up to 20 Cooper pairs, but numerical discontinuities can occur near ill-conditioned configurations and at larger sizes. The current implementation uses finite-difference approximations in the Newton--Raphson Jacobian and in the derivatives required by the Laguerre root finder. Its polynomial reconstruction solves a square system built from $M$ selected input pair energies, while the iterative procedures use update-based stopping tests. These choices keep the implementation simple and transparent, but a more systematic numerical study should assess their stability and compare them with analytic derivatives, residual-based convergence tests, and alternative root-finding strategies. The present thermodynamic analysis is restricted to selected non-degenerate spectra and to the seniority-zero sector. A systematic extension to exactly degenerate single-particle spectra, including a controlled treatment of the degenerate limit, is left for future work.

The Yang--Yang formulation discussed in subsection~\ref{YY}
suggests a complementary analytical direction. The
free-field realization of Richardson's solution within the
$\widehat{\mathfrak{su}}(2)_k$ WZW framework provides one
natural starting point for this programme
\cite{Sierra:1999mp,Sedrakyan:2010}. A related, but distinct,
connection arises from the Dotsenko--Fateev (DF) representation:
in the Gaudin setting, the saddle-point (on-shell) value of
the DF action is related to classical Virasoro conformal
blocks, while an analogous WZW on-shell quantity appears in
the CFT realization of the Richardson model and generates
the eigenvalues of its quantum integrals of motion
\cite{PNPP}. These structures suggest exploring whether
CFT and Coulomb-gas methods can provide analytical
representations of the Yang--Yang function and its critical
points, together with complementary descriptions of the
associated spectral data.

A further possibility is to investigate extensions involving
irregular conformal blocks and irregular Knizhnik--Zamolodchikov
systems~\cite{HLR,GHLR}. These topics provide
motivation for future work; no results based on irregular
conformal blocks or irregular KZ equations are claimed in
the present paper.

The source code of the Richardson solver is publicly available at
\url{https://github.com/GrzegorzBiskowski/Richardson-solver}.
Further development of the numerical implementation and the
release of the input data and scripts used to generate the
thermodynamic spectra and figures are part of our ongoing work.
These efforts are intended to improve reproducibility and
facilitate extensions to larger systems and a broader range
of non-degenerate single-particle spectra.

\section*{Acknowledgements}
This research was funded by National Science Centre, Poland 
(NCN) under grant no.~2023/49/B/ST2/01371.
	
\appendix
\section{Hard-core boson algebra}
\label{AppB1}

The pair operators
\begin{equation}
	b_j^\dagger=c_{j+}^\dagger c_{j-}^\dagger,
	\qquad
	b_j=c_{j-}c_{j+},
	\qquad
	n_j=b_j^\dagger b_j,
\end{equation}
obey the hard-core boson relations
\begin{equation}
	(b_j^\dagger)^2=b_j^2=0,
	\qquad
	[b_i,b_j^\dagger]=\delta_{ij}(1-2n_i),
	\qquad
	[n_i,b_j^\dagger]=\delta_{ij}b_i^\dagger.
	\label{app:hc}
\end{equation}
For $i\neq j$, the pair operators commute because they contain an even number of fermionic operators. For $i=j$, the first relation in \eqref{app:hc} follows from Pauli exclusion, while the commutators follow from the canonical fermionic anticommutation relations. These are the only algebraic identities needed in the Bethe ansatz derivation below.

\section{Derivation of the Richardson equations}
\label{AppB2}

For completeness, we give the short operator derivation of the Richardson equations. With
\begin{equation}
	B_\nu^\dagger=
	\sum_{j\in{\cal U}}
	\frac{b_j^\dagger}{2\epsilon_j-E_\nu},
	\qquad
	B_0^\dagger=\sum_{j\in{\cal U}}b_j^\dagger,
\end{equation}
the paired Hamiltonian takes the form
\begin{equation}
	H_{\cal U}=\sum_{j\in{\cal U}}2\epsilon_j n_j-gB_0^\dagger B_0.
\end{equation}
Using \eqref{app:hc}, one finds
\begin{align}
	[B_0^\dagger,B_\nu^\dagger]&=0,\\
	[B_0,B_\nu^\dagger]&=
	\sum_{j\in{\cal U}}
	\frac{1-2n_j}{2\epsilon_j-E_\nu},\\
	\left[H_{\cal U},B_\nu^\dagger\right]
	&=E_\nu B_\nu^\dagger
	+B_0^\dagger\left[
	1-g\sum_{j\in{\cal U}}
	\frac{1-2n_j}{2\epsilon_j-E_\nu}
	\right].
	\label{app:comm}
\end{align}
Commuting $H_{\cal U}$ through the Bethe state
\begin{equation}
	|\Psi_M\rangle=\prod_{\nu=1}^M B_\nu^\dagger|0\rangle
\end{equation}
and using $n_j|0\rangle=0$, the terms containing $n_j$ generate pairwise contributions proportional to
$(E_\mu-E_\nu)^{-1}$. After collecting all terms, one obtains
\begin{align}
	H_{\cal U}|\Psi_M\rangle
	={}&
	\left(\sum_{\nu=1}^M E_\nu\right)|\Psi_M\rangle
	\nonumber\\
	&+\sum_{\nu=1}^M
	B_0^\dagger
	\left[
	1-g\sum_{j\in{\cal U}}\frac{1}{2\epsilon_j-E_\nu}
	+2g\sum_{\mu\neq\nu}\frac{1}{E_\mu-E_\nu}
	\right]
	\prod_{\eta\neq\nu}B_\eta^\dagger|0\rangle.
	\label{app:action}
\end{align}
Thus the Bethe state is an eigenstate provided
\begin{equation}
	\frac{1}{g}
	-\sum_{j\in{\cal U}}\frac{1}{2\epsilon_j-E_\nu}
	+2\sum_{\mu\neq\nu}\frac{1}{E_\mu-E_\nu}=0,
	\qquad \nu=1,\ldots,M,
\end{equation}
which are precisely the Richardson equations~\eqref{REq}. The corresponding eigenvalue is
\begin{equation}
	E_{\mathcal U}=\sum_{\nu=1}^M E_\nu.
\end{equation}

\end{document}